\documentstyle[12pt]{article}

\begin{document}

\topmargin 0pt
\oddsidemargin 7mm
\headheight 0pt
\topskip 0mm
\addtolength{\baselineskip}{0.40\baselineskip}

\hfill SOGANG-HEP 211/97

\hfill May 1997

\vspace{1cm}

\begin{center}
{\large \bf Note on the Abelian Pure CS Theory \\
Based on the Improved BFT Method}
\end{center}

\vspace{1cm}

\begin{center}
Mu-In Park and Young-Jai Park \\
{\it Department of Physics and Basic Science Research Institute \\
Sogang University, C.P.O. Box 1142, Seoul 100-611, Korea}
\end{center}

\vspace{2cm}


We reconsider the Abelian pure Chern-Simons theory in three dimensions by using our improved Batalin-Fradkin-Tyutin
Hamiltonian formalism. As a result, we show several novel 
features, including the connection of the Dirac brackets.
In particular, through the path integral quantization, we obtain the desired new type of the Wess-Zumino action. 

\vspace{3cm}

\newpage

Recently, Banerjee [1] applied the
Batalin-Fradkin-Tyutin (BFT) Hamiltonian method [2,3]
to the second class constraint system of
the Abelian Chern-Simons (CS) field theory [4-6], which yielded a first-class-constraint algebra in an extended phase space
by introducing new fields.
As a result, he obtained a new type of Abelian 
Wess-Zumino (WZ) action [7], which could not be obtained in the usual path-integral framework.
After his work, we applied the BFT Hamiltonian method
to the self-dual massive theory[8]. As a result, we obtained the Lagrangian 
revealing both the St\"uckelberg effect [9] and the CS effect [1,10] corresponding to the first-class Hamiltonian[11].
In this analysis we also treated the pure CS theory. 
However, although that action was simpler than that of the self-dual case, we failed to find the desired full Lagrangian because an unwanted ${\delta}$ function existed in the measure of the partition function.
Furthermore, we also analyzed several other interesting examples in this approach[12].
Very recently, we improved the usual BFT formalism[13]. In particular, we have shown that this improved method is very powerful when we analyze the non-Abelian theory[14]. On the other hand, all these works [1,10-14] still do not resolve the unwanted Fourier parameter ${\xi}$ generally contained in the measure part. 

In this note, we will apply our improved BFT formalism to the pure CS theory [11] in order to solve the problems mentioned above. First, we will briefly recapitulate the usual BFT formalism.
Next, we will explain several novel features of our improved BFT method, including Dirac brackets [15] defined in the original phase space. Next, we will obtain the desired full Lagrangian corresponding to the first-class Hamiltonian. We will also show that one can directly derive this Lagrangian by using our improved BFT method when the original one is first order.

Now, we first recapitulate the usual BFT formalism by analyzing
the pure Abelian CS model
\begin{equation}
 S = \int d^3 x [ \frac{m}{2} \epsilon_{\mu\nu\rho}
         A^\mu \partial^\nu A^\rho].
\end{equation}
Since this action is invariant up to the total divergence
under the gauge transformation
$\delta A^{\mu} = \partial^{\mu} \Lambda$,
this action has an origin for the second-class constraint
which is differrent from the well-known massive Maxwell theory.
This differrnce in origin is due to the explicit 
gauge-symmetry-breaking term in the action. In fact, the origin of the second class constraints
is due to the symplectic structure of the CS model.

Following the usual Dirac standard procedure [15],
we find that there are three primary constraints,
\begin{eqnarray}
\Omega_0 &\equiv& \pi_0 \approx 0, \nonumber \\
\Omega_i &\equiv& \pi_i - \frac{1}{2} m \epsilon_{ij} A^j 
\approx 0 ~~( i = 1, 2 ),
\end{eqnarray}
and one secondary constraint,
\begin{equation}
\omega_3 \equiv  m \epsilon_{ij} \partial^i A^j \approx 0,
\end{equation}
obtained by conserving $\Omega_0$ with the total Hamiltonian,
\begin{equation}
H_T = H_c + \int d^2x [ u^0 \Omega_0 + u^i \Omega_i ],
\end{equation}
where $H_c$ is the canonical Hamiltonian,
\begin{equation}
H_c = - \int d^2x ~[  m \epsilon_{ij} A^0 \partial^i A^j ],
\end{equation}
and where $x=(t,\vec{x})$, the two-space vector
$\vec{x}=(x^1,x^2)$, $\epsilon_{12}=\epsilon^{12}=1$,
and the Lagrange multipliers are $u^0$ and $u^i$.
No further constraints are generated via this iterative procedure.
We find that all rest constraints except $\Omega_0 = \pi_0 \approx 0$
are superficially second class constraints.
However, in order to extract the true second-class constraints,
it is essential to redefine $\omega_3$ by using $\Omega_1$ and
$\Omega_2$ as follows:
\begin{eqnarray}
\Omega_3 \equiv \omega_3 + \partial^i \Omega_i =  \partial^i \pi_i + \frac{1}{2} m \epsilon_{ij} \partial^i A^j.
\end{eqnarray}
Then, $\Omega_0$ and $\Omega_3$ form the first-class algebra, while the $\Omega_i$ form the second class algebra as follows:
\begin{eqnarray}
\Delta_{i j}(x,y) \equiv \{ \Omega_{i}(x), \Omega_{j}(y) \} =
- m \epsilon_{ij}\delta^2(x-y) ~~ (i, j = 1, 2).
\end{eqnarray}

In order to convert this system into the first class one,
the first objective is to transform $\Omega_i$ into the
first-class by extending the phase space.
Following the usual BFT approach [2,3,10-12],
we introduce new auxiliary fields $\Phi^i$ to convert the 
second-class constraint $\Omega_i$ into
a first-class one in the extended phase space, and we assume that the Poisson algebra of the new fields is given by
\begin{equation}
   \{ \Phi^i(x), \Phi^j(y) \} = \omega^{ij}(x,y),
\end{equation}
where $\omega^{ij}$ is an antisymmetric matrix.
Let the original fields be $F=( A^\mu, \pi_\mu)$.
Then, the modified constraint in the extended phase space is given by
\begin{equation}
  \tilde{\Omega}_i(F, \Phi)
         =  \Omega_i + \sum_{n=1}^{\infty} \Omega_i^{(n)};
                       ~~~~~~\Omega_i^{(n)} \sim (\Phi)^n,
\end{equation}
satisfying the boundary condition
$\tilde{\Omega}_i(F; 0) = \Omega_i$.
The first-order correction term in the infinite series [3] is given by
\begin{equation}
  \Omega_i^{(1)}(x) = \int d^2 y X_{ij}(x,y)\Phi^j(y),
\end{equation}
and the first-class  constraint algebra of $\tilde{\Omega}_i$ requires the condition
\begin{equation}
   \triangle_{ij}(x,y) +
   \int d^2 w~ d^2 z~
        X_{ik}(x,w) \omega^{kl}(w,z) X_{lj}(z,y)
         = 0.
\end{equation}
As was emphasized in Ref. 1 and 10-13, there is a natural arbitrariness
in choosing $\omega^{ij}$ and $X_{ij}$ from Eq. (8) and Eq. (10),
which corresponds to the canonical transformation
in the extended phase space [2,3].
We take the simple solutions as
\begin{eqnarray}
\omega^{i j} (x,y) &=&
 \epsilon^{ij} \delta^2(x-y),      \nonumber \\
X_{i j} (x,y) &=& \sqrt{m} \delta_{ij}\delta^2(x-y),
\end{eqnarray}
which are compatible with Eq. (11), as it should be.
Using Eqs. (9), (10), and (12), the new set of constraints is found to be
\begin{equation}
\tilde{\Omega}_i = \pi_i - \frac{1}{2} m \epsilon_{ij} A^j + \sqrt{m} \Phi^i ~~~~~~~(i=1,2),
\end{equation}
which are strongly involutive,
\begin{equation}
\{ \tilde{\Omega}_{\alpha}, \tilde{\Omega}_{\beta} \} = 0~~~~~~
 (\alpha, \beta = 0, 1, 2, 3)
\end{equation}
$\mbox{with}~~\tilde{\Omega}_0 \equiv
\Omega_0~~\mbox{and}~~\tilde{\Omega}_3
\equiv \Omega_3$.
As a result, we get all the first-class constraints
in the extended phase space
by applying the BFT formalism systematically.
Furthermore, we observe that only $\Omega_i^{(1)}$ contributes
to the series, Eq. (9), defining the first-class constraint, while all higher-order terms given by Eq. (9) vanish
as a consequence of the choice in Eq. (12). Then, one can also derive the corresponding involutive Hamiltonian
in the extended phase space by using the infinite series [3]:
\begin{equation}
 \tilde{H} = H_c + \sum_{n=1}^{\infty} H^{(n)}, ~~~~~H^{(n)} \sim (\Phi)^n,
\end{equation}
satisfying the initial condition
$\tilde{H}(F; 0) = H_c$. As a result, in a previous work [11], one of us already obtained the total corresponding canonical Hamiltonian as follows
\begin{equation}
\tilde H = H_c + H^{(1)},
\end{equation}
where
\begin{equation}
H^{(1)} = \int d^2x
                  [- \sqrt{m} (\partial_i \Phi^i) A^0].
\end{equation}

However, we now would like to use our improved BFT formalism in order to find ${\tilde H}$ directly from $H_c$. To this end, first let us define the ${\it physical }$ variables $\tilde F= (\widetilde{A}^{\mu}, \widetilde{\pi}_{\mu})$ corresponding to the original ones, $F$, within the Abelian conversion in the extended phase space,
which are strongly involutive, i.e.,
\begin{eqnarray}
\{ \widetilde{\Omega}_{i}, \widetilde{F} \} =0.
\end{eqnarray}
These variables can be generally found as
\begin{eqnarray}
\widetilde{F}(F; \Phi ) &=& F + \sum^{\infty}_{n=1} \widetilde{F},~~~~~~~ \widetilde{F}^{(n)} \sim (\Phi)^{n}   \end{eqnarray}
satisfying the boundary conditions $\widetilde{F}(F; 0 ) = F$. Here, the first order iteration terms are given by
\begin{eqnarray}
\widetilde{A}^{\mu (1)}&=&
                   - \Phi^{j}\omega_{jk} X^{kl}
                   \{ \Omega_{l}, A^{\mu} \}_{(A, \pi)}  \nonumber  \\
              &=& (0,  \frac{1}{\sqrt{m}}\epsilon^{ik} \Phi^{k} ),
                              \nonumber  \\
\widetilde{\pi}_{\mu}^{(1)} &=& - \Phi^{j}\omega_{jk} X^{kl}
                   \{ \Omega_{l}, \pi_{\mu} \}_{(A, \pi)}  \nonumber \\
              &=&  ( 0, \frac{\sqrt{m}}{2} \Phi^{i}).
\end{eqnarray}
Furthermore, since the modified variables up to the first iterations, $F+\widetilde{F}^{(1)}$, are found to be involutive,
i.e., to satisfy Eq. (18), the higher order iteration terms
\begin{eqnarray}
\widetilde{F}^{(n+1)} &=&                              -\frac{1}{n+1} \Phi^{j}\omega_{jk} X^{kl} (G)^{(n)}_l      \end{eqnarray}
with
\begin{eqnarray}
(G)^{(n)}_l &=& \sum^{n}_{m=0} \{ \Omega_{i}^{(n-m)}, \widetilde{F}^{(m)}\}_{F}  + \sum^{n-2}_{m=0} 
\{ \Omega_{i}^{(n-m)}, \widetilde{F}^{(m+2)}\}_{(\Phi)}           +  \{ \Omega_{i}^{(n+1)}, \widetilde{F}^{(1)} \}_{(\Phi)}       
\end{eqnarray}
are not needed and are also found to be automatically vanishing. Hence, the physical variables $\tilde{F}$ in the extended phase space are given by

\begin{eqnarray}
&&\tilde{A}^{\mu}=(A^{0}, ~A^{i}+ \frac{1}{\sqrt{m}} 
\epsilon^{ik} \Phi^{k} ),  \nonumber \\
&&\tilde{\pi}_{\mu}=(\pi_{0}, ~\pi_i + ~\frac{\sqrt{m}}{2}
\Phi^{i} ).
\end{eqnarray}

Similar to the physical variables, $\widetilde{F}$, all other physical quantities, including the Hamiltonian, which correspond to the functions of $F$, can be also found in principle by considering solutions as in Eq. (19) [10-14].
In other words, using the elegant property [13-14] 
\begin{eqnarray}
\widetilde{K}(F; \Phi)= K(\widetilde{F})
\end{eqnarray}
for the arbitrary function or the functional $K$ defined on the original phase-space variables unless $K$ has time derivatives, the relation
\begin{eqnarray}
\{ K(\widetilde{F}),\widetilde{\Omega}_{i} \}=0
\end{eqnarray}
is generally satisfied for any function $K$ not having time derivatives because the $\widetilde{F}$ and their spatial derivatives already commute with $\widetilde{\Omega}_{i}$ at equal times by definition. 

Now, we can directly obtain the desired  first class Hamiltonian $\widetilde{H_{c}}$ corresponding
to the canonical Hamiltonian $H_{c}$ of Eq. (5) as follows:
\begin{eqnarray}
\widetilde{H_{c}}(F; \Phi) &=& H_{c}(\widetilde{F})\nonumber \\    &=& H_{c}(F) + \int d^2x [- \sqrt{m} A^0(\partial_i\Phi^i)],
\end{eqnarray}
which is involutive with the first-class constraints
\begin{eqnarray}
\{\tilde{\Omega}_i, \tilde H\} &=& 0,~~~~~~~ ( i = 1, 2, 3), \nonumber \\
\{\tilde{\Omega}_0, \tilde H\} &=& \tilde{\Omega}_3 - 
\partial^i \tilde{\Omega}_i.
\end{eqnarray}
This completes the operatorial conversion of the original
second-class system with Hamiltonian $H_c$ and the constraints
$\Omega_i$ into a first-class one with the involutive Hamiltonian $\tilde H$
and the constraints $\tilde{\Omega}_i$.
Furthermore, we can naturally generate the first-class
Gauss law constraint $\tilde{\Omega}_3$ from the time evolution
of $\tilde{\Omega}_0$ by introducing another equivalent 
Hamiltonian given by adding the trivial term 
$(\partial^i{A^0})\tilde{\Omega}_i$ to $\widetilde{H_{c}}$.
On the other hand, all the constraints already satisfy the property in Eq. (25) because $\widetilde{\Omega}_{i}(F; \Phi) = \Omega_{i}(\widetilde{F})$. In this way, the second-class constraint system $\Omega_{i}(F) \approx 0$ is converted into a first-class constraint one $\widetilde{\Omega}_{i}(F; \Phi ) =0$ with the boundary conditions $\widetilde{\Omega}_{i} |_{\Phi=0}=\Omega_{i}$.
In addition, due to this useful property,
we can also obtain the first-class Lagrangian, which will be treated later,
simply by replacing ${F}$ with $\tilde{F}$ as follows
\begin{eqnarray}
\widetilde{\cal L}&=&\frac{m}{2} \epsilon^{\mu \nu \rho}\tilde{A_{\mu}}\partial_{\nu}\tilde{A_{\rho}} \nonumber \\
             &=& {\cal L} + {\Phi}^j \left(\sqrt{m} F_{0j}  - \frac{1}{2}\epsilon_{jk}\dot{\Phi}^k 
\right),
\end{eqnarray}
where the second $\Phi$-dependent term effectively corresponds to be the WZ term in the extended phase space. 

It seems appropriate to comment on the Dirac brackets.
As is known, in the Dirac formalism [15], one can make the second class constraint system $\Omega_{i} \approx 0$ into effectively a first-class-constraint one $\Omega_{i}(F)=0$ only by deforming the phase space $F=(A^{\mu}, \pi_{\mu})$ without introducing any new field. Hence, it seems that these two formalisms are drastically different. However, remarkably the Dirac brackets can be easily read from the usual BFT-formalism by noting that the Poisson bracket in the extended phase space in the $\Phi \rightarrow 0$ limit generally becomes 
\begin{eqnarray}
\{ \widetilde{A}, \widetilde{B} \} |_{\Phi=0}
              &=&   \{A, B \} - \{A, \Omega_{k} \} \Delta^{kk'} \{\Omega_{k'}, B \}  \nonumber \\
            &=&   \{A, B \}_{D},
\end{eqnarray}
where $\Delta^{kk'}=-X^{lk} \omega_{ll'} X^{l'k'}$ is the inverse of $\Delta_{kk'}$ in Eq. (6)[13,14,16]. As a result, the non-trivial Poisson brackets of the pure CS theory in the extended phase space, which are given by
\begin{eqnarray}
&&\{\tilde{A}^{i}(x), \tilde{A}^{j}(y) \}  =\frac{1}{m} \epsilon^{ij} \delta^{2}(x-y),  \nonumber \\
&&\{\tilde{\pi}_{i}(x), \tilde{\pi}_{j}(y) \}  =\frac{m}{4} \epsilon_{ij} \delta^{2}(x-y),  \nonumber \\
&&\{\tilde{A}^{i}(x), \tilde{\pi}_{j}(y) \}  =\frac{1}{2}\delta^{i}_{j} \delta^{2}(x-y),
\end{eqnarray}
are exactly the same as the usual Dirac brackets [5].

Finally, we consider the path integral quantization
in order to obtain the quantum Lagrangian corresponding to $\tilde{H_c}$
in the Hamiltonian formalism. This analysis can be achieved by evaluating the partition function
\begin{eqnarray}
Z_0 &=& \int {\cal D} \pi_{\mu} {\cal D} A^{\mu}
           {\cal D} \Phi^{i}
           \prod^3_{\alpha, \beta = 0}
            \delta(\tilde{\Omega}_{\alpha}) \delta({\Gamma}_{\beta})
            det \mid \{\tilde{\Omega}_{\alpha}, {\Gamma}_{\beta}\} \mid
            e^{\frac{i}{\hbar}S_0}, \nonumber \\
S_0 &=& \int  d^3 x ~( \pi_{\mu} {\dot A^{\mu}} 
            + \Phi^{2} {\dot \Phi^{1}}
            -  \tilde {\cal H}_c ),
\end{eqnarray}
according to the Faddeev-Popov(FP) formula [17], where we regard the new fields $( \Phi^{1}, \Phi^{2} )$
as conjugate pairs [1,10-14]. In the usual evaluation of the partition function in Eq. (31), an unwanted $\xi$ field, which is introduced to
exponentiate the Gauss law constraint $\tilde{\Omega}_3$, can not be removed in the determinent of measure when we make the action covariant[1,10-14]. However, we will show that the covariant action can be obtained without this drawback by the
method adopted by Fujiwara et al. [18].

The first step is to introduce the gauge fixing condition
\begin{eqnarray}
\Gamma_0 =A_0 \approx 0
\end{eqnarray}
for the primary constraint $\tilde{\Omega}_0 =\pi_0 \approx 0$ and to perform the
path integral over $A_0, \pi_0$ such that Eq. (31) reduces to
\begin{eqnarray}
Z_I &=& \int {\cal D} \pi_{i} {\cal D} A^{i}
           {\cal D} \Phi^{i}
           \prod^2_{j=1}
            \delta(\tilde{\Omega}_{j}) \delta(\tilde{\Omega}_{3}) \prod^3_{K,L=1}\delta({\Gamma}_{K})^{'}
            det \mid \{\tilde{\Omega}_{K}, {\Gamma}_{L}\}^{'} \mid
            e^{\frac{i}{\hbar}S_I},\nonumber \\
S_I &=& \int  d^3 x ~\left( \pi_{i} {\dot A^{i}} 
            + \Phi^{2} {\dot \Phi^{1}}
              \right),
\end{eqnarray}
where we have used the delta function $\delta(\tilde{\Omega _0})$ and the result,
\begin{eqnarray}
            det \mid \{\tilde{\Omega}_{\alpha}, {\Gamma}_{\beta}\} \mid
            = N \cdot  det \mid \{\tilde{\Omega}_{K}, {\Gamma}_{L}\} \mid ~~~~~~~ (K,~L=1,2,3)
\end{eqnarray}
with some number $N$, will eventually be infinite.
However, that does not present the usual problem because by
explicitly manipulating $\{\tilde{\Omega}_{\alpha}, \Gamma_0 \}$ for the gauge
function (32) the determinant of the $4 \times 4$ matrix is reduced to
that of a $3 \times 3$ matrix without the constraints $\tilde{\Omega}_0$ and
corresponding gauge function $\Gamma_0$ sector. Moreover, the primed parts in the measure denote that the conditions $\tilde{\Omega}_0=\pi_0 \approx 0$ and $\Gamma_0=A_0 \approx 0$ are imposed after the calculations. 
The second step is to represent the delta function of the constraints
containing the derivatives of the momentum, i.e., the Gauss law constraint
$\delta(\tilde{\Omega_3})$ by the $\xi$ integration
$\int {\cal D} \xi e^{(i/\hbar) \int \xi \tilde{\Theta}_3 }$ and to perform the
path integration over $\pi_i$ by exploiting the delta function
$\delta(\tilde{\Omega_i})$. Then, the partition function 
in Eq.(33) is reduced to be
\begin{eqnarray}
Z_{II} &=& \int {\cal D} A^{i}
           {\cal D} \Phi^{i}
           \prod^3_{K,L=1}\delta({\Gamma}_{K})^{''}
            det \mid \{\tilde{\Omega}_{K},{\Gamma}_{L} \}^{''} \mid
            e^{\frac{i}{\hbar}S_{II}},      \nonumber \\ 
S_{II} &=& \int  d^3 x ~\left( \frac{m}{2}\partial^0 A^{i}             \epsilon_{ij}A^j + m \xi \epsilon_{ij} \partial^i A^j 
- \sqrt{m} \Phi^i(\partial^0 A^i -\partial^i \xi)
             + \Phi^{2} {\dot \Phi^{1}} \right),
\end{eqnarray}
where the doubly primed parts in the measure factor denote that the CS constraints $\tilde{\Omega}_i$ which are imposed after all the manipulations. Of course, for FP-type gauges [17], which do not involve the momenta, this additional restriction is useless. However, we maintain this restriction for the generality of our treatment.

Now, we get the covariant form, except for the auxilliary-field part, $\Phi^i$, of
the partition function, by identifying the auxilliary field $\xi$ by the $A^0$ field as follows:
\begin{eqnarray}
Z_{III} &=& \int {\cal D} A^{\mu}
             {\cal D} \Phi^{1} {\cal D} \Phi^{2}
           \prod^3_{K,L=1}\delta({\Gamma}_{K})^{''}
            det \mid \{\tilde{\Omega}_{K}, {\Gamma}_{L}\}^{''}\mid
            e^{\frac{i}{\hbar}S_{III}}, \nonumber \\
S_{III} &=& \int  d^3 x ~\left(
\frac{m}{2} \epsilon^{\mu \nu \rho}{A_{\mu}}\partial_{\nu}{A_{\rho}}
- {\Phi}^2 (\sqrt{m} F^{02} - \dot{\Phi}^1 ) 
- {\Phi}^1 (\sqrt{m} F^{01}  ) \right).
\end{eqnarray}
Note that the action in Eq. (36) is exactly the same as the result in Eq. (28) which was derived by
the simple replacement of $F$ with $\tilde F$ in the original action up to total divergence. Here, the momentum integration for the field $\Phi^i$ remains intact, but
the momentum integration of all the other fields are carried out to produce the Lagrangian path integral. 

Now, let us perform the path integreation over $\Phi^2$ by considering the $\Phi^2$-independent gauge function 
$\Gamma_k$. As a result, we obtain
\begin{eqnarray}
Z_{IV} &=& \int {\cal D} A^{\mu}
             {\cal D} \Phi^{1}
           \prod^3_{k,l=2}\delta({\Gamma}_{k})^{''}
            det \mid \{\tilde{\Omega}_{k}, {\Gamma}_{l}\}{''}\mid
  \delta \left[ \sqrt{m} F^{02} -(D_0 \Phi^1)  \right]
            e^{\frac{i}{\hbar}S_{IV}}, \nonumber \\
S_{IV} &=& \int  d^3 x ~ \left( \frac{m}{2} \epsilon_{\mu \nu \rho}~A_{\mu} \partial_{\nu} A_{\rho}
            - {\Phi}^{1} {\sqrt m} F^{01} \right).
\end{eqnarray}
Here, the delta function 
$\delta \left[ \sqrt{m} F^{02} - {\partial}_0 \Phi^1 \right]$ is the unwanted term in the measure part. 

However, if we adopt the procedure i) Fourier transformation, i.e.,
\begin{eqnarray}
\delta \left[ \sqrt{m} F^{02} -{\partial}_0 \Phi^1  \right] 
 =\int {\cal D} \xi exp\{-\frac{i}{\hbar} \xi
  (\sqrt{m} F^{02} -{\partial}_0 \Phi^1)\},
\end{eqnarray}
ii) removal of $A^1$ with the gauge fixing $\Gamma_1=A^1 \approx 0$, and iii) recovery of $A^1$ by redefining 
$\xi$ as $\xi \equiv \sqrt{m} A^{1} $, then the final partition function can be expressed as
\begin{eqnarray}
Z_{F} &=& \int {\cal D} A^{\mu}
             {\cal D} \Phi^{1}
           \prod^3_{k,l=2}\delta({\Gamma}_{k})'''
            det \mid \{\tilde{\Omega}_{k}, {\Gamma}_{l}\}''' \mid
            e^{\frac{i}{\hbar}S_{F}}, \nonumber \\
S_{F} &=& \int  d^3 x ~ \frac{m}{2} \epsilon_{\mu \nu \rho}
            ~A_{\mu} \partial_{\nu} A_{\rho}
                      +S_{WZ_F}, \nonumber \\
S_{WZ_F}&=& - \int d^3 x  {\sqrt m} \left(  {\Phi}^{1} F^{01} \right),
\end{eqnarray}
where the triply primed parts in the measure denote that the conditions $\Gamma_1=A_1 \approx 0$, $\tilde{\Omega _i} \approx 0, and~ \Omega_0 \approx 0$ are imposed after all the manipulations. Note that when we take the unitary gauge $\Gamma_2= {\Phi}^{1} \approx 0$,
the original theory is easily reproduced.

In summary, we have shown that our improved BFT method,
which more effectively converts a second-class system into a 
first-class one, is very powerful for finding all the
desired physical quantities defined in the extended phase space 
and for finding the first class Hamiltonian by re-analyzing the pure CS theory[11], which has a
different origin, a second-class structure. 
We have also shown that the Poisson brackets of the modified physical fields $\tilde F$
in the extended phase space are just the usual Dirac brackets of the fields $F$ defined in the original phase space.
Furthermore, through path-integral quantization, we obtained the desired new WZ action which solved the problems of the unwanted $\delta$-function and the Fourier parameter $\xi$
in the measure part by using the removal and recovery techniques.

Finally we would like to comment that our WZ action is different in that the WZ term does not affect the gauge symmetry of the original action with the extended gauge-symmetry transformation $A_{\mu} \rightarrow A_{\mu} + \partial_{\mu} \Lambda,~ \Phi^1 \rightarrow \Phi^1$. Hence, the only role of the new WZ action is to make the second-class system into the corresponding first-class one.

\vspace{1cm}

\begin{center}
{\bf ACKNOWLEDGMENT}
\end{center}

The present study was supported by
the Basic Science Research Institute Program,
Ministry of Education, 1997, Project No. BSRI-97-2414.

\end{document}